\documentclass{optica-article}

\journal{opticajournal} 

\articletype{Research Article}

\usepackage{mhchem}
\usepackage{lineno}
\usepackage{bm}
\usepackage{siunitx}
\usepackage{upgreek}
\usepackage[title]{appendix}
\usepackage{float}

\begin{document}

\title{Symmetric silicon microring resonator optical crossbar array for accelerated inference and training in deep learning}

\author{Rui Tang,\authormark{1,4} Shuhei Ohno,\authormark{1} Ken Tanizawa,\authormark{2} Kazuhiro Ikeda,\authormark{3} Makoto Okano,\authormark{3} Kasidit Toprasertpong,\authormark{1} Shinichi Takagi,\authormark{1} and Mitsuru Takenaka\authormark{1,5}}

\address{\authormark{1}Department of Electrical Engineering and Information Systems, The University of Tokyo, Tokyo 113-8656, Japan\\
\authormark{2}Quantum ICT Research Institute, Tamagawa University, Tokyo 194-8610, Japan\\
\authormark{3}National Institute of Advanced Industrial Science and Technology, Ibaraki 305-8568, Japan}

\email{\authormark{4}ruitang@mosfet.t.u-tokyo.ac.jp\\
\authormark{5}takenaka@mosfet.t.u-tokyo.ac.jp}


\begin{abstract*} 
Photonic integrated circuits are emerging as a promising platform for accelerating matrix multiplications in deep learning, leveraging the inherent parallel nature of light. Although various schemes have been proposed and demonstrated to realize such photonic matrix accelerators, the in-situ training of artificial neural networks using photonic accelerators remains challenging due to the difficulty of direct on-chip backpropagation on a photonic chip. In this work, we propose a silicon microring resonator (MRR) optical crossbar array with a symmetric structure that allows for simple on-chip backpropagation, potentially enabling the acceleration of both the inference and training phases of deep learning. We demonstrate a $4 \times 4$ circuit on a Si-on-insulator (SOI) platform and use it to perform inference tasks of a simple neural network for classifying Iris flowers, achieving a classification accuracy of 93.3\%. Subsequently, we train the neural network using simulated on-chip backpropagation and achieve an accuracy of 91.1\% in the same inference task after training. Furthermore, we simulate a convolutional neural network (CNN) for handwritten digit recognition, using a $9 \times 9$ MRR crossbar array to perform the convolution operations. This work contributes to the realization of compact and energy-efficient photonic accelerators for deep learning.
\end{abstract*}

\section{Introduction}
Deep learning has ignited transformative breakthroughs across diverse fields. At its core, a deep learning system relies on an artificial neural network (ANN) comprising potentially more than billions of artificial neurons. These neurons are meticulously arranged into multiple layers. As input data traverse these layers, they undergo a sequence of linear and nonlinear operations. The primary linear operation of an ANN during both its inference and training phases is matrix multiplication, a computationally intensive process that consumes significant time and energy. This necessity has spurred the development of high-performance and energy-efficient hardware dedicated to deep learning.

Photonic integrated circuits (PICs) are emerging as a promising platform for accelerating matrix multiplications in deep learning, leveraging the inherent parallel nature of light \cite{shen2017deep, wetzstein2020inference, zhou2022photonic, mcmahon2023physics, peserico2023integrated}. Various schemes have been proposed and demonstrated to realize such photonic matrix accelerators \cite{clements2016optimal, tait2016microring, tait2017neuromorphic, feldmann2021parallel, xiao2021large, tang2021ten, tang2022two, ohno2022si, zhang2022silicon, moralis2022neuromorphic, sludds2022delocalized, bandyopadhyay2022single, giamougiannis2023coherent, giamougiannis2023analog, dong2023higher, tang2024lower}. Despite significant progress, the in-situ training of ANNs using photonic matrix accelerators remains challenging due to the difficulty of direct on-chip backpropagation on a photonic chip \cite{spall2023training, zheng2023dual}. For the architecture based on Mach-Zehnder interferometer (MZI) meshes, an on-chip backpropagation method was proposed by Hughes et al. and recently demonstrated by Pai et al \cite{hughes2018training, pai2023experimentally}. However, this method requires the use of transparent on-chip photodetectors (PDs) which are not commonly available in existing foundries, or a bulky monitoring camera which hinders integration. In contrast, the optical crossbar array based on microring resonators (MRRs) is a promising architecture that allows for direct on-chip backpropagation \cite{ohno2022si}. Two separate input ports in this optical crossbar array facilitate the simple injection of optical signals in both forward and backward directions. However, the structure proposed in our previous work is asymmetric, leading to unbalanced insertion losses among all optical paths \cite{ohno2022si}. This asymmetry distorts the implemented matrices and impedes the scalability of the circuit.

In this work, we propose a novel MRR crossbar array with a symmetric structure, eliminating unbalanced insertion loss among all optical paths and thereby addressing the issues identified in our previous work. We demonstrate a 4$\times$4 circuit on a Si-on-insulator (SOI) platform and use it to perform inference tasks of a simple neural network for classifying Iris flowers, achieving a classification accuracy of 93.3\%. Subsequently, we train the neural network using simulated on-chip backpropagation and achieve an accuracy of 91.1\% in the same inference task after training. Furthermore, we simulate a convolutional neural network (CNN) for handwritten digit recognition, using a $9 \times 9$ MRR crossbar array to perform the convolution operations.

\section{Structure and device}

\begin{figure}[b]
\centering\includegraphics[width=13cm]{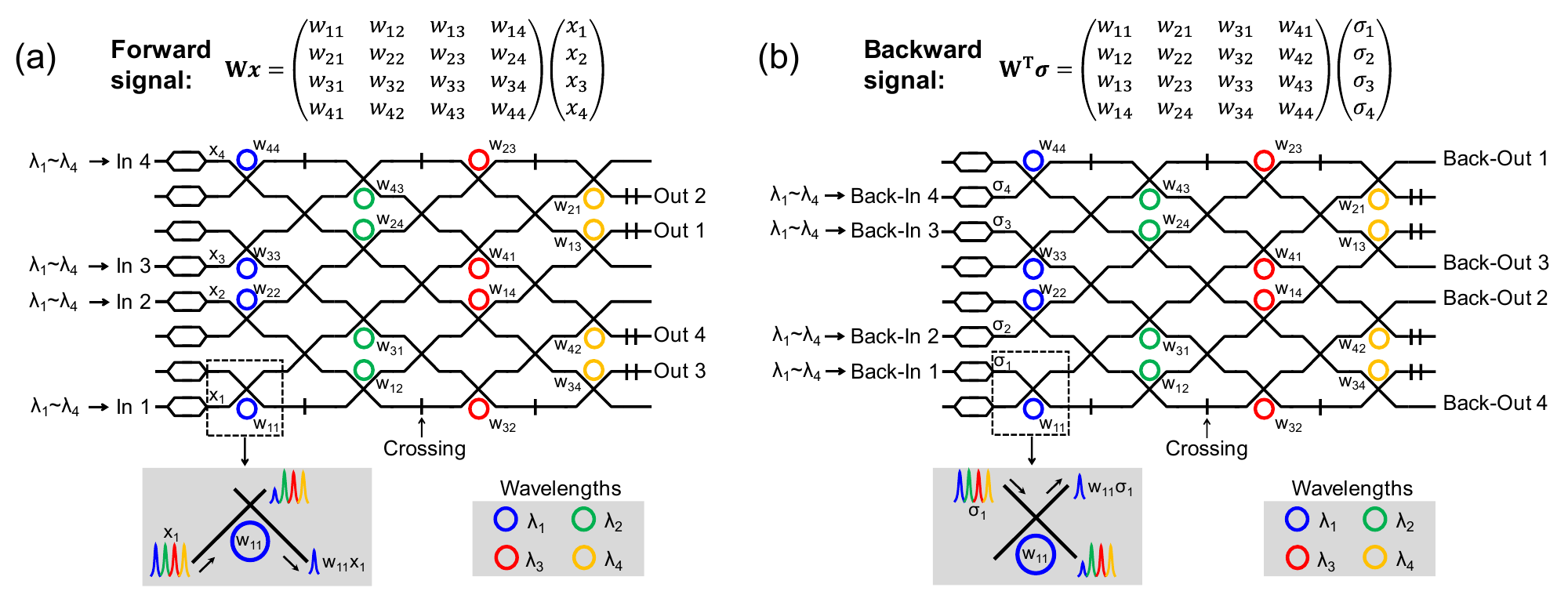}
\caption{Proposed optical crossbar array. The matrix and vector are generated by MRRs and MZIs, respectively. Multiple wavelengths are injected into 4 input ports simultaneously. The MRRs are tuned to align with different wavelengths, and the associated matrix element is represented by the transmittance of optical power at the drop port. (a) By injecting a forward signal $\boldsymbol x$, which represents the output signal from the previous layer in an ANN, the crossbar array performs the multiplication between $\mathrm{\bf W}$ and $\boldsymbol x$. (b) By injecting a backward signal $\boldsymbol \sigma$, which represents the error signal backpropagated from the next layer in an ANN, the crossbar array performs the multiplication between $\mathrm{\bf W^\top}$ (the transpose of $\mathrm{\bf W}$) and $\boldsymbol \sigma$.}
\end{figure}

The proposed symmetric MRR crossbar array is schematically illustrated in Fig. 1. For $N \times N$ matrices ($N = 4$ in Fig. 1), this structure uses $N^2$ MRRs to represent a matrix and $2N$ Mach-Zehnder interferometers (MZIs) to generate two input vectors: $\boldsymbol x$ and $\boldsymbol \sigma$ (referred to as the forward and backward signals, respectively). Note that $\boldsymbol x$ and $\boldsymbol \sigma$ are not injected into the circuit at the same time. The forward signal $\boldsymbol x$, representing the output signal from the previous layer in an ANN, is multiplied by the weight matrix $\mathrm{\bf W}$, as shown in Fig. 1(a). Here, $N$ wavelengths are injected into the $N$ input ports simultaneously. Each MRR is tuned to couple with one wavelength, and the associated matrix element is represented by the transmittance of optical power at the drop port. At each output port for the forward signal, $N$ optical signals coupled through different MRRs are multiplexed into the same waveguide and detected by an on-chip or external PD. Therefore, the multiplication and accumulation operations are performed at the MRRs and the PDs, respectively. Similarly, the backward signal $\boldsymbol \sigma$, representing the error signal backpropagated from the next layer in an ANN, is multiplied by the transposed weight matrix $\mathrm{\bf W^\top}$ without reconfiguring the MRRs, as shown in Fig. 1(b). This automatically performs on-chip backpropagation in a simplified way, different from MZI-based schemes. Here, different input and output ports are used compared with Fig. 1(a). In addition, two extra waveguide crossings are inserted into the output port for $\boldsymbol x$ to balance the insertion loss, because the light at the drop port of an MRR does not pass through the waveguide crossing in the forward direction but passes through the waveguide crossing twice in the backward direction, as illustrated by the MRR for the matrix element $w_{11}$ in Figs. 1(a) and 1(b). In our previous work \cite{ohno2022si}, the number of waveguide crossings in various optical paths with unequal lengths ranged from 0 to $2N-2$ for the forward signal and from 0 to $2N$ for the backward signal, due to its asymmetric structure. In contrast, each optical path in this new structure has $2N$ waveguide crossings for both forward and backward signals. Therefore, by employing this symmetric circuit topology, all optical paths by design have uniform lengths and insertion losses, effectively addressing the issues identified in our previous work.

\begin{figure}[t]
\centering\includegraphics[width=11.5cm]{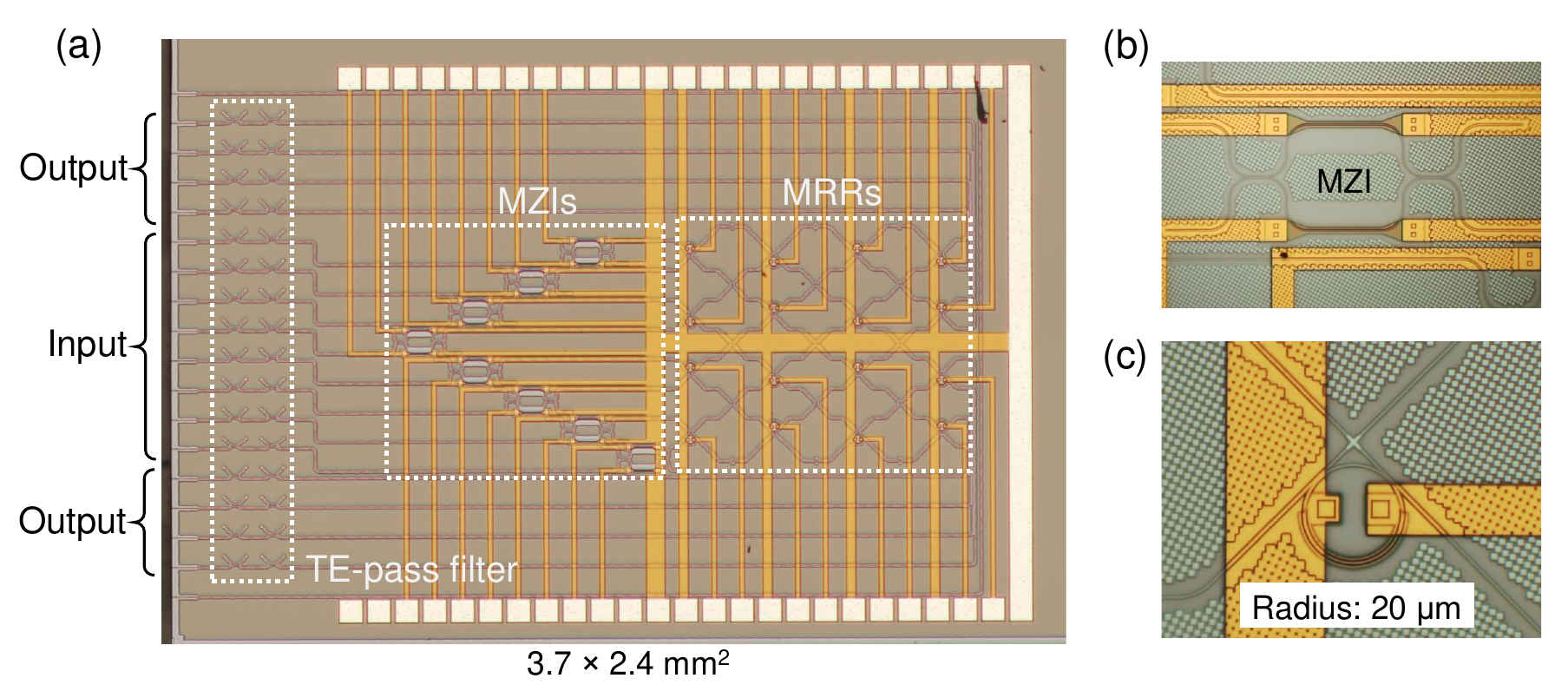}
\caption{Microscope images of a $4 \times 4$ MRR crossbar array fabricated on a SOI platform. (a) The entire circuit consists of MZIs, MRRs, and TE-pass filters. The consumed chip area is 3.7 $\times$ 2.4 \unit{\mm^2}. (b) Enlarged view of one MZI. Only one input port and one output port are used. The other two ports are terminated with inverse waveguide tapers. (c) Enlarged view of one MRR. The radii of all MRRs are 20 \unit{\micro\meter}.}
\end{figure}

Figure 2(a) shows a compact $4 \times 4$ MRR crossbar array fabricated on an SOI platform with a chip area of 3.7 $\times$ 2.4 \unit{\mm^2}, and Figs. 2(b) and 2(c) show the enlarged views of one MZI and one MRR, respectively. The single-mode waveguide is 440 nm wide and 220 nm high, and the propagation loss of the fundamental transverse electric (TE) mode is 1.3 dB/cm. The radii of all MRRs are 20 \unit{\micro\meter}, corresponding to a free spectral range (FSR) of 4.4 nm. Polarization filters based on directional couplers are employed at all input and output ports to filter out residual transverse magnetic (TM) light. Waveguide crossings with low insertion loss and low crosstalk are utilized \cite{ma2013ultralow}. In this $4 \times 4$ circuit, we omitted the two waveguide crossings at the output ports for the forward signal, as they would simply introduce the same amount of optical loss to the four output ports. Such uniform loss can be easily compensated for. Thermo-optic phase shifters based on thin metal heaters are employed to tune the MZIs and MRRs, with a power consumption of approximately 19.3 mW/$\uppi$. In the future, we anticipate that the adoption of ultralow-power electro-optic phase shifters can significantly reduce power consumption and eliminate thermal crosstalk, enabling a more compact and energy-efficient circuit \cite{ohno2021si, wakita2023low, wakita2024add}. 

\section{Results}

\subsection{Experimental setup}
Figure 3 shows the experimental setup and the packaged chip. The chip is wire-bonded for external electrical control and packaged with a fiber array for stable fiber coupling, with an average fiber-chip coupling loss of approximately 2.9 dB/facet. Four continuous-wave (CW) lights at different wavelengths ($\uplambda_1$--$\uplambda_4$: 1549.00, 1549.75, 1550.50, 1551.25 nm) are generated by a 4-channel tunable laser (Agilent, N7714A) and combined into a single optical fiber by inversely using two stages of $1 \times 2$ optical splitters. The combined lights are injected into a micro-electromechanical systems (MEMS) optical switch, which directs light to the ports for either the forward or backward signal. Polarization controllers are used to adjust the polarizations of input light to the TE. A temperature controller is used to stabilize the chip temperature at room temperature. The phase shifters of MZIs and MRRs are controlled by a 40-channel programmable direct current (DC) power supply (Nicslab Ops, XPOW). The intensities of output optical signals are measured by a 24-channel high-speed optical power meter (OptoTest Corp., OP760). A computer sends commands to the DC driver and fetches data from the optical power meter via Python codes.

\begin{figure}[t]
\centering\includegraphics[width=10cm]{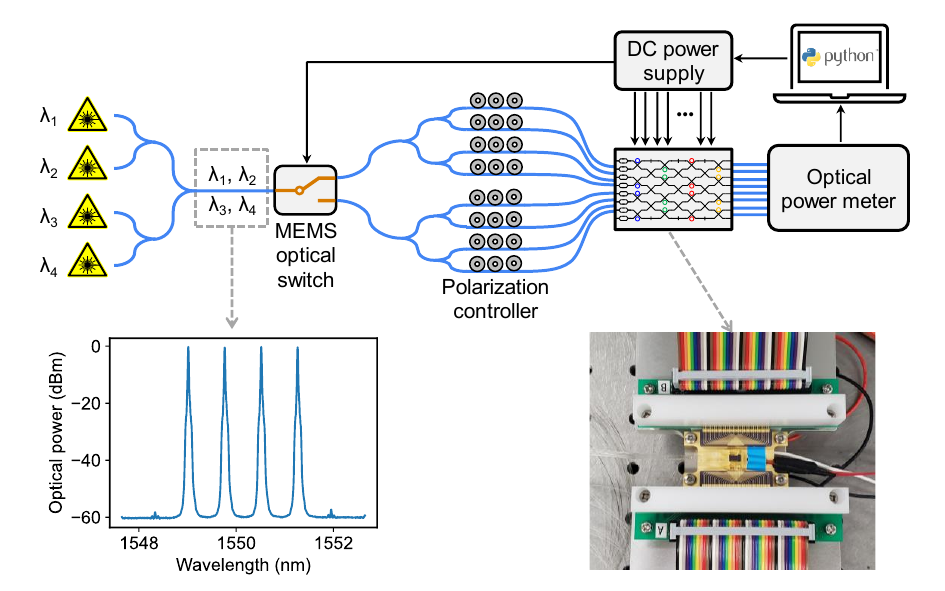}
\caption{Experimental setup. Four CW lights at different wavelengths are generated by a 4-channel tunable laser and combined into a single optical fiber by inversely using two stages of $1 \times 2$ optical splitters. The MEMS optical switch directs the combined light to the ports for either the forward or backward signal. The chip is wire-bonded for external electrical control and packaged with a fiber array for stable fiber coupling.}
\end{figure}

\subsection{Device characterization}
Each MZI is characterized by sweeping the electric power applied to the phase shifter on one MZI arm and measuring the optical power response at the output port. The MRRs are tuned to have negligible light coupled into them during the characterization of MZIs. The transmission of the MZI at the In 1 port as a function of heater power is shown in Fig. 4(a), exhibiting a high extinction ratio of 51 dB. The worst extinction ratio among all MZIs is 37.6 dB, and random initial phases have been observed when no electric power is applied to the heaters of MZIs. These results are provided in Appendix A. Next, we sequentially select each input port and measure the transmission spectrum at associated output ports to characterize the MRRs. No electric power is applied to the phase shifters for MRRs during their characterizations. Figure 4(b) shows the transmission spectra measured at the Out 1-4 ports when light is injected into the In 1 port. The resonant wavelengths of the four MRRs slightly differ due to fabrication non-uniformity. In the future, low-loss phase shifters based on phase change materials (PCMs) can be integrated on the MRRs to compensate for initial phase differences \cite{fang2022ultra, miyatake2023proposal}. The measured FSRs are 4.4 nm, consistent with the design value. The insertion loss of each MRR at resonant wavelengths near 1550 nm is approximately 1.9 dB. The characterization results of all MRRs are provided in Appendix A. To characterize the difference between the forward and backward directions, for each MRR, we inject light at one wavelength into the two input ports of the MRR, respectively, then sweep the heater power and measure the optical power response at the two output ports, as illustrated in Fig. 4(c). Figure 4(d) shows the results of characterizing the two directions of the MRR for the matrix element $w_{12}$. It can be observed that the two directions exhibit almost the same characteristics. However, due to the non-uniform coupling loss between fibers and edge couplers, the forward and backward paths of a few MRRs show relatively large differences. These results are provided in Appendix A.

\begin{figure}[t]
\centering\includegraphics[width=11cm]{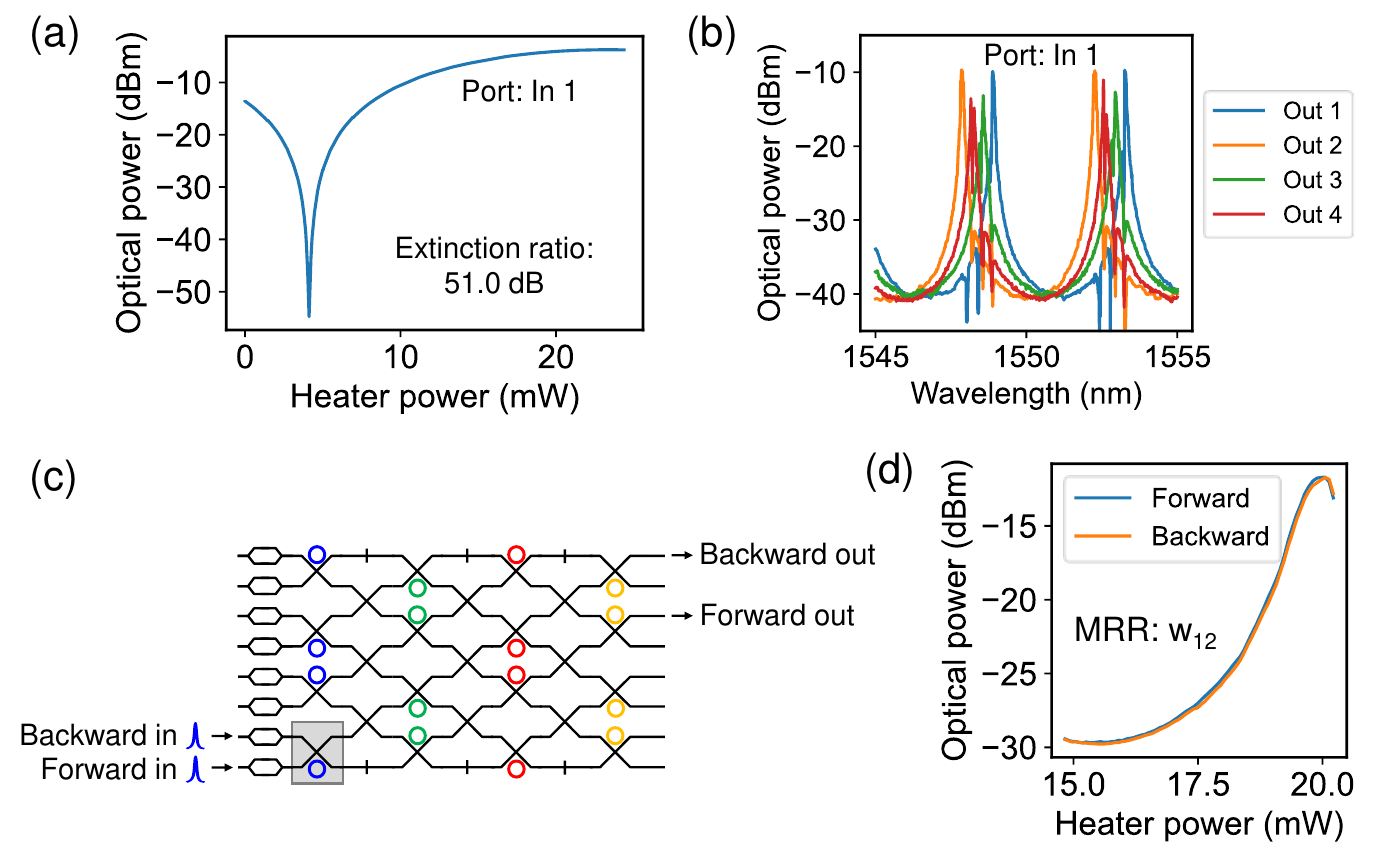}
\caption{Characterizations of MZIs and MRRs. (a) Characterization result of the MZI in the In 1 port as a function of heater power. The MZI exhibits a high extinction ratio of 51 dB. (b) Transmission spectra measured at the Out 1-4 ports when sweeping the wavelength of light injected into the In 1 port. No electric power is applied to the phase shifters for MRRs. The resonant wavelengths of the four MRRs slightly differ due to fabrication non-uniformity. (c) Illustration of characterizing the difference between the forward and backward paths of each MRR. (d) Optical power measured at the output ports for forward and backward signals of one MRR. The two directions exhibit almost the same characteristics.}
\end{figure}

\subsection{Matrix implementation}
After characterizing all MZIs and MRRs, we implement various matrices by controlling the heaters of MRRs. The results are shown in Fig. 5. Each matrix element is measured by setting one MZI into the maximum-transmittance state and the others into the minimum-transmittance state. The heater power for each MRR is calibrated so that the maximum optical power coupled via each MRR is almost the same. It can be seen that the matrices measured from the forward and backward directions are transposed to each other. In contrast to the result in our previous work [13], where the matrices for the forward and backward directions exhibited significant differences, here the desired matrices are successfully realized for both directions with negligible differences. Error signals in these matrices are suppressed to the level of approximately -15 dB.

\begin{figure}[t]
\centering\includegraphics[width=13cm]{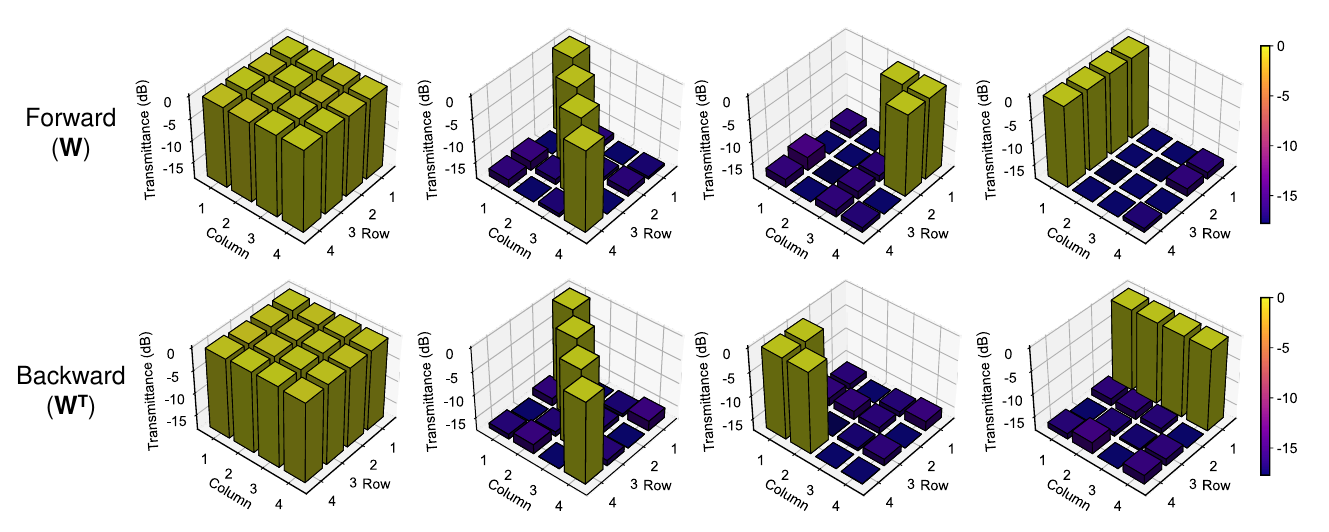}
\caption{Experimental implementations of various matrices for forward and backward signals. Each matrix element is measured by setting one MZI into the maximum-transmittance state and the others into the minimum-transmittance state. The matrices measured from the forward and backward directions are transposed to each other. Error signals in these matrices are suppressed to the level of approximately -15 dB.}

\centering\includegraphics[width=7cm]{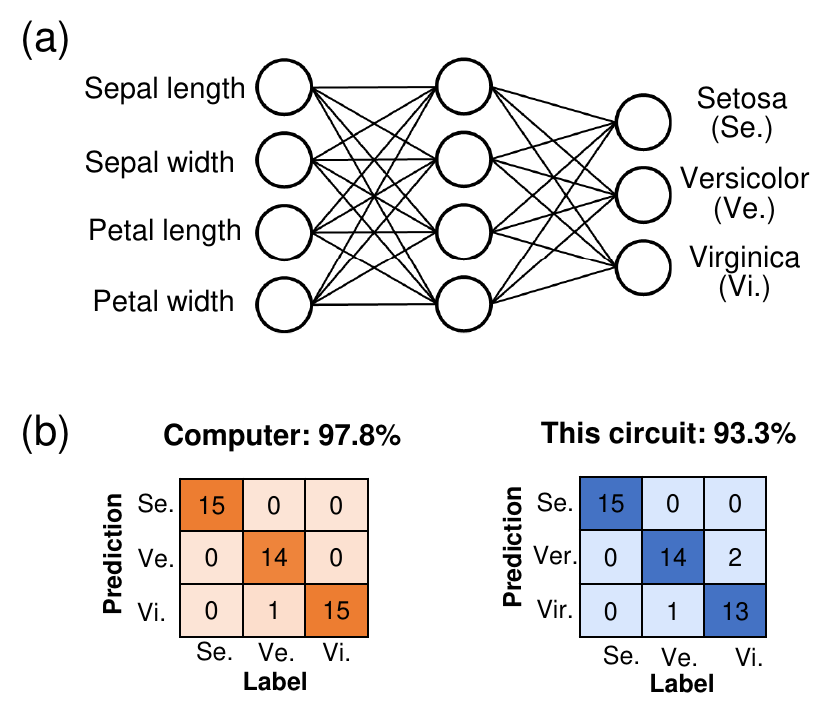}
\caption{Inference tasks using the optical crossbar array. (a) A 3-layered neural network for classifying Iris flowers. The sigmoid function is used as the nonlinear activation function. (b) Inference results after the neural network is trained on a computer using the stochastic gradient descent algorithm. Classification accuracies of 97.8\% and 93.3\% are obtained using the computer and this circuit, respectively.}
\end{figure}

\subsection{Inference task}
We construct a 3-layered neural network for classifying Iris flowers, as shown in Fig. 6(a). The network takes a 4-element vector as input, which includes sepal length/width and petal length/width, and generates a 3-element vector as output, representing the flower species. The sigmoid function is used as the nonlinear activation function. The dataset consists of a total of 150 samples, 105 of which are used for training and the remaining 45 for testing \cite{iris}. All data are normalized between 0 and 1 for physical implementations on the circuit. We first trained this network on a computer using the stochastic gradient descent algorithm and achieved 97.8\% classification accuracy using the test data on the computer. Then, we normalize the weight matrices and use the MRR crossbar array to perform the matrix-vector multiplications. Only forward signals are needed in the inference task. The nonlinear activation function is still performed on the computer. For the same test data, a high classification accuracy of 93.3\% is obtained, as shown in Fig. 6(b). During the experiments, we observed relatively large fluctuations in the output optical powers, which may be caused by an insufficiently stable temperature control of MRRs. We performed time-averaging measurements to reduce the noise in the experiments. This issue may be solved by using electro-optic phase shifters that do not generate heat or applying feedback controls on the thermo-optic phase shifters \cite{tait2018feedback, huang2020demonstration, wakita2024add}. Unlike state-of-the-art analog processors with ultrahigh processing speed or energy efficiency \cite{chen2023all, xu2024large}, the computation speed of this proof-of-concept circuit is limited by its small scale and slow thermo-optic phase shifters, resulting in a speed of approximately $3.2 \times 10^5$ operations per second under a clock frequency of 100 kHz. In the future, the speed can be significantly improved by scaling up the circuit and using metal-oxide-semiconductor (MOS) phase shifters \cite{wakita2024add}.

\subsection{Training with simulated on-chip backpropagation}
\begin{figure}[t]
\centering\includegraphics[width=13cm]{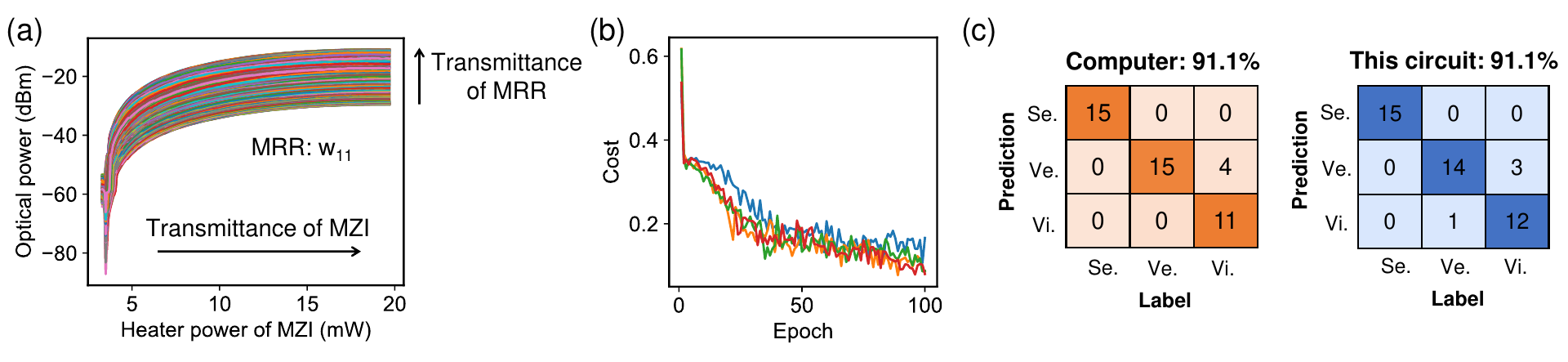}
\caption{Training of the neural network using simulated on-chip backpropagation. (a) A look-up table for one MRR characterized in the forward direction that maps the settings of an MZI and the MRR to the measured output power. (b) Changes in cost functions in four independent trainings, all of which converged successfully after 100 epochs. (c) Inference results after training using the simulated on-chip backpropagation. Classification accuracies of 91.1\% are obtained using both the computer and this circuit.}
\end{figure}

Due to the power fluctuation mentioned above, no training with direct on-chip backpropagation is performed since it would require a significant amount of measurement time. Instead, we create look-up tables for all MRRs that map the settings of MZIs and MRRs to the measured output powers. One such look-up table for an MRR characterized in the forward direction is shown Fig. 7(a). The heater power of the MZI in the In 1 port is increased from 3.3 to 19.7 mW, and the heater power of the MRR for matrix element $w_{11}$ is increased from 22.1 to 27.8 mW. Using these look-up tables, we can obtain the multiplication results between each vector element and each matrix element after simple normalizations. Since both forward and backward signals are needed in the training, for a few MRRs that exhibit relatively large differences between the forward and backward directions due to non-uniform fiber coupling loss, we compensate for this difference by adding a constant bias term to the port with lower power. We then train the neural network using the stochastic gradient descent method and simulated on-chip backpropagation. The flowchart of on-chip backpropagation is given in our previous work \cite{ohno2022si}. The mean squared error is used as the cost function during training. In each iteration, the multiplications between matrix and vector elements are performed by fetching data from the look-up tables, and all other operations are performed by the computer. Figure 7(b) shows the changes in cost functions in four independent trainings, all of which converged successfully after 100 epochs. After training, we normalize the obtained weight matrices and perform the inference tasks using the test data again. The results are shown in Fig. 7(c). Classification accuracies of 91.1\% are obtained using both the computer and this circuit. A slight difference between the two results is attributed to the power fluctuations in measured optical powers, which are not fully captured in the look-up tables. 

\subsection{Simulation of handwritten digit recognition}
We further simulate a CNN for handwritten digit recognition, using a $9 \times 9$ MRR crossbar array circuit as a photonic tensor core to perform the convolution operations. The network structure is illustrated in Fig. 8(a). Each input image is a single-channel image with $28 \times 28$ pixels. An input image is first processed by nine different $3 \times 3$ kernels. After applying a rectified linear unit (ReLU) function, the max pooling layer extracts the maximum elements in $2 \times 2$ windows. Then, the multi-channel data is flattened into a vector and processed by fully-connected layers. The ReLU function and the softmax function are used in the hidden layer with 100 nodes and the output layer, respectively. To perform the convolution operations using the MRR crossbar array, each $3 \times 3$ kernel is flattened into a 9-element row vector, and these nine kernels are converted into a $9 \times 9$ matrix. Since negative elements can exist in this matrix, the matrix is normalized into a non-negative matrix with elements between 0 and 1 for physical implementation, using a method described in our previous work \cite{ohno2022si}.

\begin{figure}[t]
\centering\includegraphics[width=9cm]{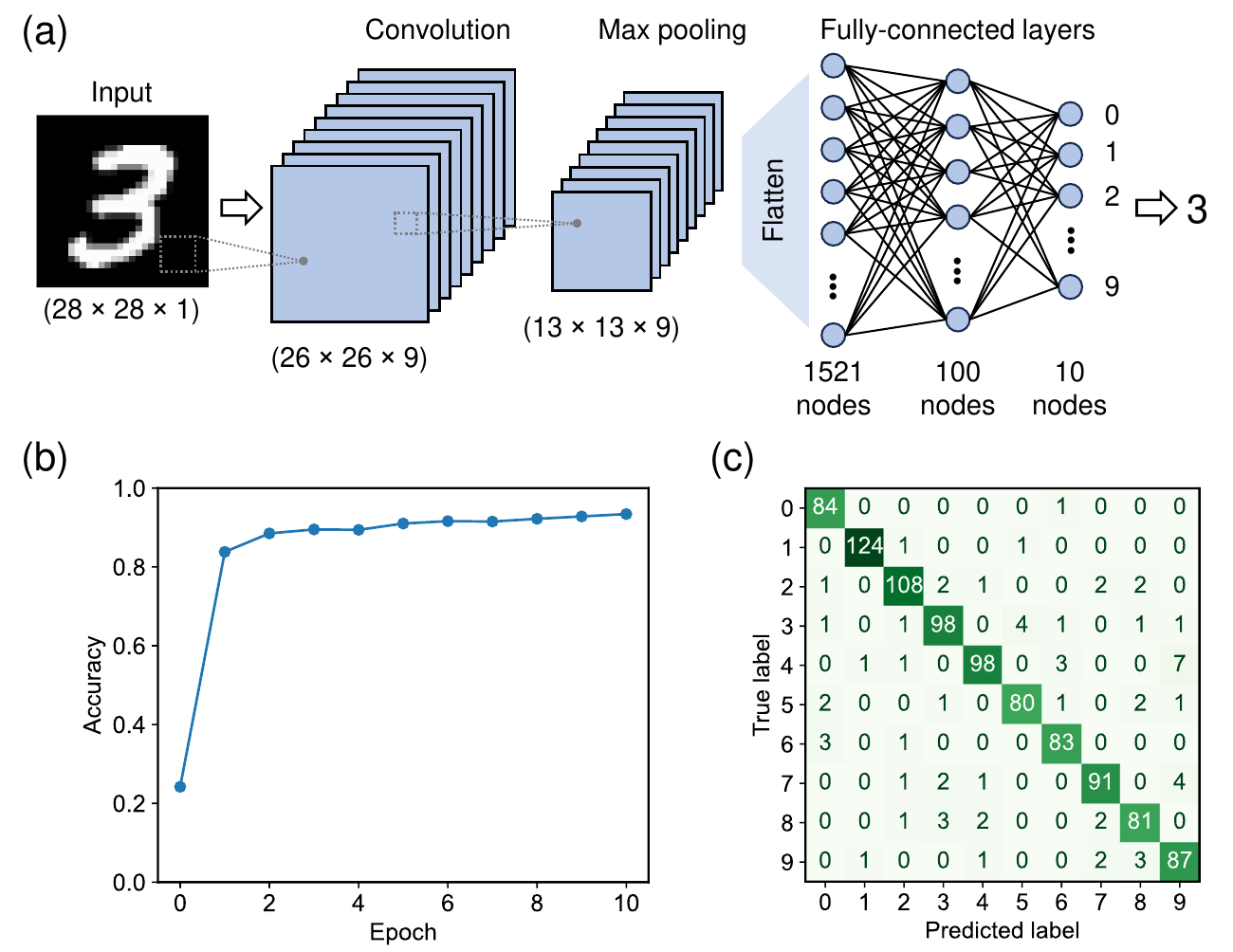}
\caption{Simulation of handwritten digit recognition using a $9 \times 9$ MRR crossbar array. (a) A CNN for handwritten digit recognition. A $9 \times 9$ MRR crossbar array is used to perform the convolution operations. (b) The changes in classification accuracies during training. The accuracy reaches 93.4\% after 10 epochs. (c) Confusion matrix after 10 epochs of training.}
\end{figure}

We developed a model to calculate the output power of each wavelength for the MRR crossbar array, taking into account the crosstalk from other wavelength channels. These crosstalks depend on the quality factors of the MRRs, and we have assumed all MRRs have quality factors of approximately $3 \times 10^5$. Assuming the same ring radii (20 \unit{\micro\meter}) used in the demonstrated circuit, the transmission spectrum at the drop port and through port of each MRR can be calculated \cite{bogaerts2012silicon}. For a $9 \times 9$ circuit, we assume that the 9 input wavelengths are evenly distributed within one FSR of the MRRs near the 1550 nm wavelength. For a given forward/backward signal, we inject the modulated multi-wavelength signals into the forward/backward ports, and then calculate the transmitted optical signals at each drop/through port stage by stage. At each output port, the optical powers of these multi-wavelength signals are summed up and normalized to obtain the result of the MVM.

We used the MNIST database to train and test this neural network. The MNIST database consists of 60000 images for training and 10000 images for testing, from which we selected the first 10000 training images and the first 1000 test images in our simulation. We use the ADAM optimizer to train the network and the $9 \times 9$ circuit to perform the backpropagation and inference in the convolution layer \cite{kingma2014adam}. Figure 8(b) shows the changes in classification accuracies on the test data during the training process. The accuracy reaches 93.4\% after 10 epochs, and the confusion matrix is shown in Fig. 8(c). Higher accuracies can be expected using more kernels and more complicated networks.

\section{Conclusion}
We have proposed and demonstrated a symmetric MRR crossbar array for accelerated inference and training in deep learning. Using a $4 \times 4$ MRR crossbar array to perform matrix-vector multiplications in a pretrained 3-layered neural network for classifying Iris flowers, we achieved a high classification accuracy of 93.3\% in the inference task. Subsequently, we trained the neural network using simulated on-chip backpropagation and achieved an accuracy of 91.1\% in the same inference task after training. Furthermore, we simulated a CNN for handwritten digit recognition, using a $9 \times 9$ MRR crossbar array to perform the convolution operations. This work contributes to the realization of compact and energy-efficient photonic accelerators for deep learning.

\begin{appendices}
\section{Device characterizations}
Figures 9 and 10 show the characterization results of all MZIs and MRRs, respectively. Most of the MZIs have high extinction ratios greater than 40 dB. The drop ports of MRRs exhibit high extinction ratios of approximately 30 dB. Figure 11 shows the results of characterizing the difference between the forward and backward paths of each MRR, as illustrated in Fig. 4(c). Only a few MRRs have relatively large differences between the forward and backward paths, due to non-uniform fiber-waveguide coupling losses.

\begin{figure}[H]
\centering\includegraphics[width=12.5cm]{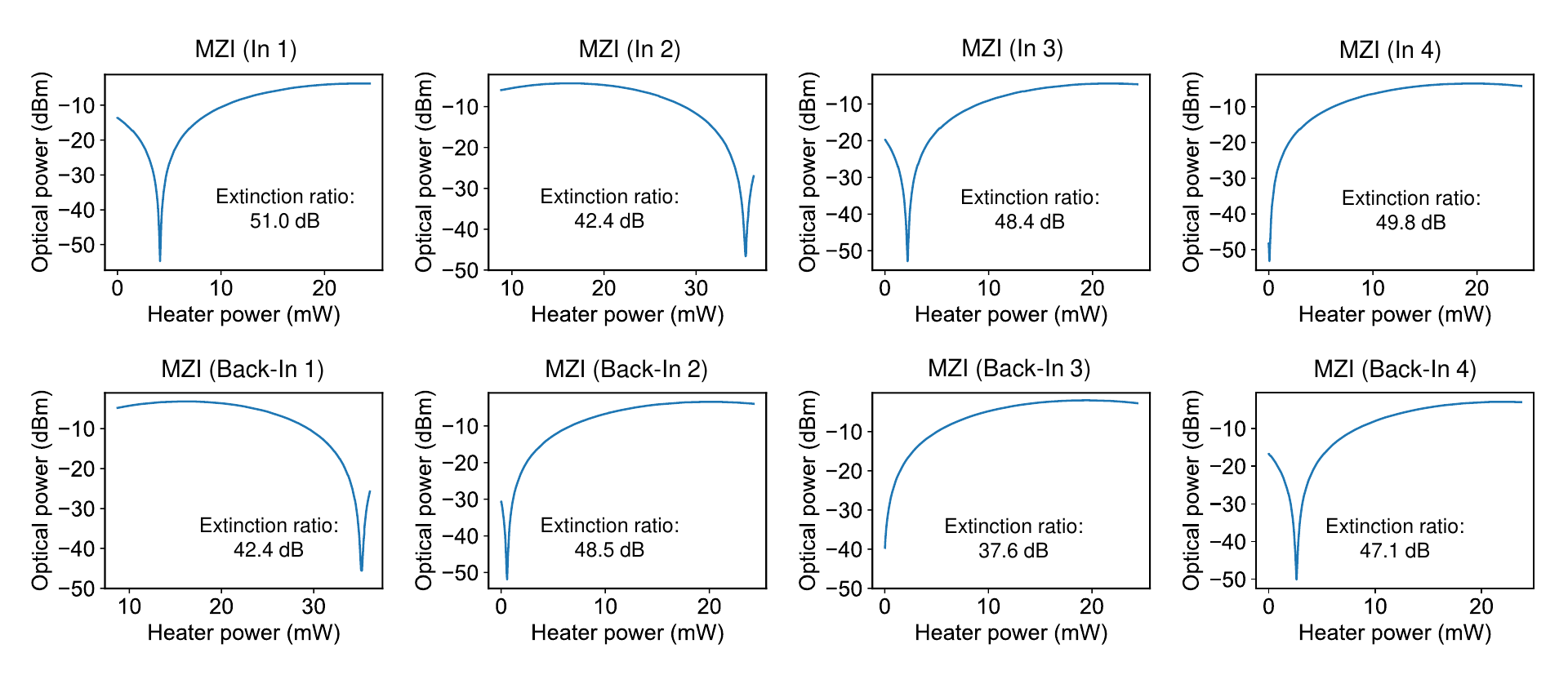}
\caption{Characterization results of all MZIs.}
\end{figure}

\begin{figure}[H]
\centering\includegraphics[width=12.5cm]{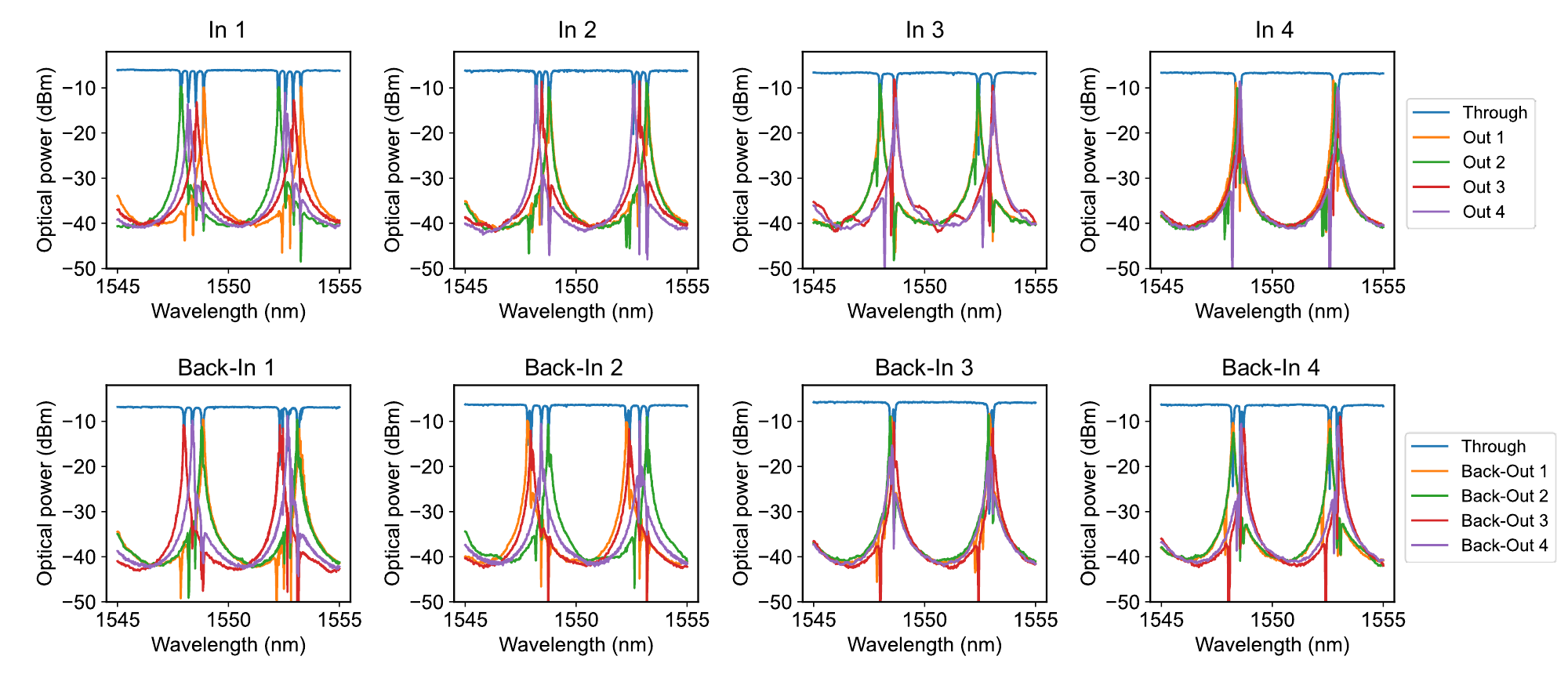}
\caption{Characterization results of all MRRs. No electric power is applied to the heaters of MRRs.}
\end{figure}

\begin{figure}[H]
\centering\includegraphics[width=12.5cm]{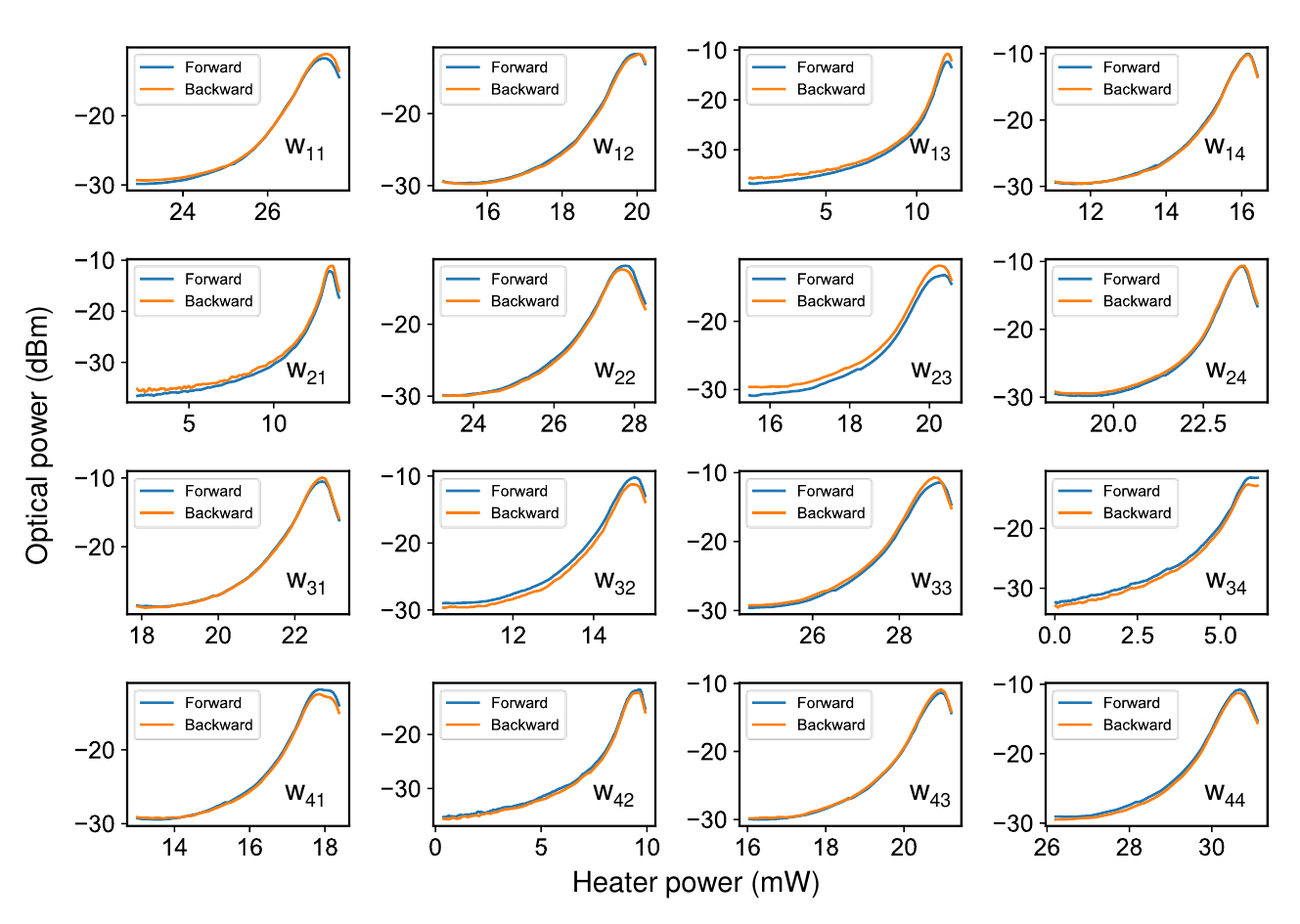}
\caption{Measured differences between the forward and backward paths of all MRRs.}
\end{figure}

\end{appendices}

\begin{backmatter}
\bmsection{Funding}
Japan Science and Technology Agency (JST) CREST (JPMJCR2004); Japan Society for the Promotion of Science (JSPS) KAKENHI (22K14298).

\bmsection{Author contributions}
S.O., K.T., and R.T. conceived the structure. S.O. designed the chip. K.I. and M.O. fabricated the chip. R.T. performed the experiments and wrote the manuscript with input from all the other authors. M.T. supervised the project.

\bmsection{Disclosures}
The authors declare no competing interests.

\bmsection{Data availability}
Data underlying the results presented in this paper are available from the corresponding authors upon reasonable request.


\end{backmatter}

\bibliography{sample}

\end{document}